\documentclass[10pt,letterpaper]{article}
\usepackage[top=0.85in,left=2.75in,footskip=0.75in]{geometry}

\usepackage{amsmath,amssymb}

\usepackage{changepage}

\usepackage[utf8x]{inputenc}

\usepackage{textcomp,marvosym}

\usepackage{cite}

\usepackage{nameref,hyperref}

\usepackage[right]{lineno}

\usepackage{microtype}
\DisableLigatures[f]{encoding = *, family = * }

\usepackage[table]{xcolor}

\usepackage{array}

\newcolumntype{+}{!{\vrule width 2pt}}

\newlength\savedwidth

\newcommand\thickhline{\noalign{\global\savedwidth\arrayrulewidth\global\arrayrulewidth 2pt}%
\hline
\noalign{\global\arrayrulewidth\savedwidth}}

\raggedright
\usepackage[aboveskip=1pt,labelfont=bf,labelsep=period,justification=raggedright,singlelinecheck=off]{caption}

\makeatletter
\renewcommand{\@biblabel}[1]{\quad#1.}
\makeatother

\usepackage{lastpage,fancyhdr,graphicx}
\usepackage{epstopdf}
\fancyheadoffset[L]{2.25in}
\fancyfootoffset[L]{2.25in}
\usepackage{subcaption}

\usepackage{xspace}
\usepackage{soul}
\usepackage{cancel}
\usepackage{mathtools}
\usepackage{makecell}
\usepackage{nicefrac}
\usepackage{float}
\usepackage{wrapfig}
\usepackage{xcolor}

\definecolor{dgreen}{rgb}{0.0, 0.5, 0.0}

\usepackage{textcomp}

\usepackage{siunitx}
\usepackage{longtable}
\usepackage{multirow}
\usepackage{cleveref}

\crefname{equation}{}{}

\begin{document}
\vspace*{0.2in}

\begin{flushleft}
{\Large
\textbf\newline{A comprehensive dynamic growth and development model of \textit{Hermetia illucens} larvae} 
}
\newline
\\
Murali Padmanabha,
Alexander Kobelski,
Arne-Jens Hempel,
Stefan Streif\textsuperscript{*}\\
\bigskip
Automatic Control and System Dynamics Lab, Technische Universitt Chemnitz, Chemnitz, 09107, Germany
\\
\bigskip

%
%





* stefan.streif@etit.tu-chemnitz.de

\end{flushleft}
\graphicspath{{./gfx/pdf/}}

\section*{Abstract}
Larvae of \textit{Hermetia illucens}, also commonly known as black soldier fly (BSF) have gained significant importance in the feed industry, primarily used as feed for aquaculture and other livestock farming.
Mathematical model such as Von Bertalanffy growth model and dynamic energy budget models are available for modelling the growth of various organisms but have their demerits for their application to the growth and development of BSF.
Also, such dynamic models were not yet applied to the growth of the BSF larvae despite models proven to be useful for automation of industrial production process (e.g. feeding, heating/cooling, ventilation, harvesting, etc.).
This work primarily focuses on developing a model based on the principles of the afore mentioned models from literature that can provide accurate mathematical description of the dry mass changes throughout the life cycle and the transition of development phases of the larvae.
To further improve the accuracy of these models, various factors affecting the growth and development such as temperature, feed quality, feeding rate, moisture content in feed, and airflow rate are developed and integrated into the dynamic growth model.
An extensive set of data were aggregated from various literature and used for the model development, parameter estimation and validation.
Models describing the environmental factors were individually validated based on the data sets collected.
In addition, the dynamic growth model was also validated for dry mass evolution and development stage transition of larvae reared on different substrate feeding rates.
The developed models with the estimated parameters performed well highlighting its application in decision-support systems and automation for large scale production.

\section{Introduction}
\textit{Hermetia illucens}, commonly known as the black soldier fly (BSF), is an insect species which is widely studied for the high nutrition value of its larvae. 
Studies \cite{Newton1977,Barragan-Fonseca2017,Wang2017,Smetana2019}, showcase these nutritional values and its suitability as a source for animal feed and human food. 
Several studies, \cite{Newton2005} and \cite{Nguyen2015,Salomone2017,Bava2019,Spranghers2017} amongst the recent, also indicate their application for recycling food and bio waste. 
These studies clearly demonstrate the potential of \textit{Hermetia illucens} in addressing the approaching food scarcity while reducing the resource usage for their production.	
Irrespective of the potential applications of \textit{Hermetia illucens}, for their (mass) production, it is necessary to study: 
(1) the underlying biological processes such as assimilation, respiration, morphological changes, etc.; 
(2) the fundamental resource prerequisites such as feed composition, growing environment conditions, etc.;
(3) the resulting growth dynamics that exhibits the various stages of the larval growth in response to the supplied resources; and
(4) the interaction between the larvae and its environment (microbiome, substrate, etc.) and  biological effects that trigger certain events (e.g. fleeing from substrate due to low $O_2$ concentration etc.).

This insect species originates from tropical South American climate zones and thus requires warm and humid environment. 
Such conditions were verified in the research studies: 
\cite{Chia2018} highlighted the threshold temperatures and thermal requirements; 
\cite{Shumo2019} compared the development rates over different temperature ranges; 
and \cite{Holmes2012} studied the effect of humidity on the egg eclosion and adult emergence.	
The influence of diet, its moisture content and the temperatures were studied together to showcase its importance in the development of the larvae \cite{Diener2009,Harnden2016,Guo-Hui2014,Gligorescu2018}.
Another study \cite{Meneguz2018}, proposed and showed the effects of pH levels of the substrate (feed), in which larvae are grown, on the larval development.	 
From these studies, one can conclude the importance of the environmental conditions (temperature, humidity, etc.), the substrate conditions (moisture, pH, etc.) and the feed composition for the growth and development of the larvae.

A thorough literature survey revealed only time-invariant static models that describe certain biological processes of the BSF larvae.
The authors of \cite{Chia2018} suggested a model to describe the development rate as a function of temperature and, similarly, a model for calculation of metabolic rate as a function of temperature was presented in \cite{Gligorescu2019}.
In case of \cite{Palma2018}, a logistic model was suggested for modelling the larval growth in response to the air flow rate.
Also, a more recent work \cite{Sripontan2019} suggested the use of a Richards model to fit the larval growth.
These models from literature are mostly static models and do not adequately describe various time-dependent dynamical aspects of larvae production such as resource dynamics, environment dynamics, etc.
Also, it can be observed that the motivation behind the above mentioned literature was to improve the growth and hence the large scale production of BSF larvae.
In order to fully utilize such models for performing simulation studies, reactor design, process design, automation, control and resource optimization, it is also necessary to appropriately formulate them as dynamic models.
The main aim of this work is to develop a suitable mathematical model that adequately describe the effect of environmental conditions on the larval metabolism; the larval growth describing the evolution of its dry mass over time; and finally, the transition of development stages between larvae and pupae.

The following sections provide a detailed approach taken to develop the models, analyze the data and obtain the model parameters.
Firstly, a detailed explanation of the experimental setup is provided.
Then, a dynamic model describing the growth and development of the larvae is presented.
The experiments performed for the estimation of parameters are described followed by the model parameter estimation. 
Finally, the results of the models are compared with the actual measurement data and the quality of fit is determined for the models.

\section{Materials and methods}
In this work, data for model development, parameter estimation and model validation are mainly obtained from literature and experiments performed in this study.
Details of the experiments performed and the data source are also provided.
The following sections provide the details of the mathematical models developed in this work and the procedure followed for estimating the model parameters.

\subsection{Production unit}
The studies on the production of larvae in an artificial controlled environment in this work are conducted in a custom built production unit \cite{Padmanabha2019} that can provide the necessary growing condition and simultaneously perform measurements of various parameters (e.g. air and substrate temperature, CO$_2$ and O$_2$ concentrations, humidity).
The controlled environment has a volume of \SI{75}{\liter} and holds a growing tray of dimension \SI{22}{\centi\meter} x \SI{32}{\centi\meter} x \SI{5.5}{\centi\meter} that could contain up-to \SI{4}{\kilogram} of substrate.
This growing tray serves as a container for the growing medium that contains selected feed for the larvae, selected number of young larvae (neonates) and the microbiome that eventually develops and grows along with the larvae in the growing medium.
The temperature, humidity, airflow/air-concentration, and day-night cycles/photo period within the unit can be regulated as required.
Information related to the states inside the production unit such as temperature of air and growing medium; CO$_2$ and O$_2$ concentrations; and humidity in air and moisture in growing medium are recorded by the sensors integrated within. Similarly, information related to the states outside the production unit, e.g., temperature, humidity and CO$_2$ concentration of external air source, are logged using data loggers. Further details regarding the production environment can be found in \cite{Padmanabha2019} (see Section 2.1.4 and 3.7). 

\subsection{Experiment setup for moisture dependency}
To study the dependency of substrate or feed moisture, a larval growth experiment was performed.
In this experiment, ten small containers of height \SI{8}{\centi\metre} and diameter \SI{5}{\centi\metre} were filled with \SI{10}{\gram} of dry feed and varying amounts of water, from 0 to \SI{40}{\gram}, were added.
Then, 20 larvae of about 8 days old with a starting weight of \SI{2}{\milli\gram} were added to each container.
A small net was placed over each container to prevent the larvae from escaping while allowing air exchange.
All containers were then placed inside the production unit with air temperature set at 29\si{\celsius} and air ventilation at a rate of \SI{7.5}{\litre\per\minute}.
The weight of each container was checked daily.
Any changes in container weight, considered mostly due to evaporation, was supplemented to keep the moisture constant.
The final fresh and dry weight of the larvae (dried for 6~\si{\hour} at 70\si{\celsius} in an air dryer) was measured at the end of the experiment (on 8th day).

\subsection{Modelling approach for larvae growth, development and the influence of its environment}
The main focus of this work is to obtain a model that not only describes the evolution of dry mass over a given period in response to various environmental factors but also in addition to capture the drop in larval dry mass due to the maturation process that BSF larvae undergo during their last larval instars.
Furthermore, it is also necessary to obtain information related to the development phases of the larvae that can be used for streamlining the production process.
This description can be assistive in determining harvesting strategies such as harvesting for maximum larval dry weight or for obtaining pupae for rearing adults. 

The most commonly used models to describe the growth of biological organism, among others, are the von Bertalanffy growth model \cite{Bertalanffy1957} and dynamic energy budget (DEB) \cite{Kooijman2008} model. 
The former model describes the growth empirically while the latter is based on mechanistic description using the concepts of energy reserves and volume.
Despite having a simple model structure, the von Bertalanffy model can be used to model the dry weight/size change over time. However, no inference can be obtained regarding the current development phase of the larvae or the drop in dry weight during maturation.
The DEB model in comparison, uses states (energy density and structural volume) that are either difficult or not directly measurable.
Also, it is not evident if using this model, information pertaining to the development phases could be obtained.
Therefore, in this work, a new model is developed based on the mass balance approach and uses concepts such as asymptotic maximum size proposed for use with von Bertalanffy model \cite{Banavar2002} and the concept of maturity reserves used in DEB model.
The following section provides some background to the fundamentals of larval growth and an overview of the model development based on these fundamental principles.
Table~\ref{table:syms_list} lists all the symbols used in this work for developing the models.
{
	\setlength{\LTleft}{-2.25in}
	\setlength{\LTright}{0in}
	\setlength{\LTcapwidth}{\dimexpr\textwidth+4.5in\relax}
	\begin{longtable}{@{\extracolsep{\fill}}lll@{}}
		\caption{\bf List of symbols used in the description of the models.} \label{table:syms_list} \\ 
		\hline	\bf	Symbol                     & \bf Description                                		& \bf Unit                  \\ \thickhline \endfirsthead
		\caption[]{(continued ...)} \\
		\hline	\bf	Symbol                     & \bf Description                                		& \bf Unit                  \\ \thickhline 	\endhead
		\multicolumn{3}{r}{{continued on next page ...}} \\ \thickhline	\endfoot
		\hline \hline \endlastfoot
		$B_\mathrm{dry}$           & dry mass per larva 				        & [\si{\gram}]  \\
		$B_\mathrm{wet}$           & wet mass per larva 				        & [\si{\gram}]  \\
		$B_\mathrm{eff}$           & non structural assimilates in larva        & [\si{\gram}]  \\
		$B_\mathrm{str}$           & structural mass of the larva body 	        & [\si{\gram}]  \\
		$T_\mathrm{\Sigma}$        & development sums of larvae from neonates to prepupa    			& [\si{\hour}]  \\
		$B_\mathrm{feed}$          & total feed (dry mass) available in the growing medium & [\si{\gram}]  \\		
		$T_\mathrm{med}$           & temperature of growing medium in production unit   & [\si{\celsius}]                \\
		$W_\mathrm{med}$           & total water in the growing medium          & [\si{\kilogram}]               \\
		$W_\mathrm{med\%}$         & moisture concentration of substrate & [\si{\kilogram\per\kilogram}]               \\
		$C_\mathrm{air}$           & CO$_2$ concentration of air in production unit 	& [\si{\kilogram\per\meter\cubed}]  \\
		$O_\mathrm{air}$           & O$_2$ concentration of air in production unit 		& [\si{\kilogram\per\meter\cubed}]  \\
		$H_\mathrm{air}$           & absolute humidity of the air in production unit    & [\si{\kilogram\per\meter\cubed}]  \\
		$A_\mathrm{air}$           & air flow rate to the larvae production unit	& [\si{\litre\per\minute}]  \\
		$\phi_\mathrm{B_{ing}}$  & flux of feed from substrate into the larva 	& [\si{\gram\per\second}]    \\
		$\phi_\mathrm{B_{excr}}$ & flux of non digested feed back to substrate 	& [\si{\gram\per\second}]    \\
		$\phi_\mathrm{B_{assim}}$  & feed converted into energy and spent to digest the ingested feed  	& [\si{\gram\per\second}]    \\
		$\phi_\mathrm{B_{mat}}$  & assimilates spent towards building of new structure 	& [\si{\gram\per\second}]    \\
		$\phi_\mathrm{B_{maint}}$  & assimilates spent for maintenance of existing structure 	& [\si{\gram\per\second}]    \\
		$\phi_\mathrm{B_{eff}}$  & effective assimilates available from the ingested feed for growth and maintenance 	& [\si{\gram\per\second}]    \\
		$\phi_\mathrm{B_{metab}}$  & total assimilates spent for metabolic activities 	& [\si{\gram\per\second}]    \\
		$k_\mathrm{\alpha_{excr}}$  & fraction of ingested feed excreted out 	& [-]    \\
		$k_\mathrm{\alpha_{assim}}$  & fraction of ingested spent for digestion	& [-]    \\
		$\epsilon_\mathrm{inges}$  & efficiency of the ingested feed	& [-]    \\
		$k_\mathrm{inges}$  	& specific ingestion rate of larva	& [\si{\gram\per\gram\per\second}]    \\
		$k_\mathrm{maint}$  	& specific rate of maintenance and maturity of larva	& [\si{\gram\per\gram\per\second}]    \\
		$k_\mathrm{dev_{ts}}$  	& conversion factor to obtain development sums in hours & [\si{\per\second}]   \\
		$k_\mathrm{T_{\Sigma}1}$  	& development point at which the assimilation process starts to cease & [\si{\hour}]   \\
		$k_\mathrm{T_{\Sigma}2}$  	& development point at which the assimilation process ends & [\si{\hour}]   \\
		$k_\mathrm{T_{\Sigma}3}$  	& development point at which the larval development phase ends & [\si{\hour}]   \\
		$k_\mathrm{B_{asy}}$  	& asymptotic size of the larvae in dry mass & [\si{\gram}]    \\
		$k_\mathrm{T_{L}}$  	& lower boundary temperature for Arrhenius equation & [\si{\kelvin}]    \\
		$k_\mathrm{T_{ref}}$  	& reference temperature for Arrhenius equation & [\si{\kelvin}]    \\
		$k_\mathrm{T_{H}}$  	& upper boundary temperature for Arrhenius equation & [\si{\kelvin}]    \\
		$k_\mathrm{T_{AL}}$  	& Arrhenius temperature for the lower boundary temperature $k_\mathrm{T_L}$ & [\si{\kelvin}]    \\
		$k_\mathrm{T_{A}}$  	& Arrhenius temperature for the reference temperature $k_\mathrm{T_{ref}}$ & [\si{\kelvin}]    \\
		$k_\mathrm{T_{AH}}$  	& Arrhenius temperature for the upper boundary temperature $k_\mathrm{T_H}$ & [\si{\kelvin}]    \\
		$k_\mathrm{r_{max}T}$  	& maximum observed development rate in response to temperature (Logan-10 model) & [\si{\per\second}]   \\
		$k_\mathrm{r_{base}T}$  & minimum development rate observed above the lower temperature boundary (Logan-10 model) & [\si{\per\second}]   \\
		$k_\mathrm{r_{\rho}T}$  & development rate change per degree change in temperature (Logan-10 model) & [\si{\per\celsius}]   \\
		$k_\mathrm{T_{base}}$   & lower temperature boundary above which the development is observed (Logan-10 model)& [\si{\celsius}]   \\
		$k_\mathrm{T_{max}}$  	& lethal maximum temperature for larval survival (modified Logan-10 model)& [\si{\celsius}]   \\
		$k_\mathrm{\delta T}$  	& width of the high temperature boundary (modified Logan-10 model)& [\si{\celsius}]   \\
		$k_\mathrm{r_{max}dm}$  	& maximum development rate of the larvae in response to feed density/availability & [\si{\per\second}]   \\
		$k_\mathrm{B_{half}dm}$ & feed density/feeding rate fow which the development rate is half & [\si{\gram\per\gram\per\day}]   \\
		$k_\mathrm{r_{max}gm}$  	& maximum growth rate of the larvae in response to feed density/availability & [\si{\gram\per\second}]   \\
		$k_\mathrm{B_{half}gm}$ & feed density/feeding rate for which the development rate is half & [\si{\gram\per\gram\per\day}]   \\
		$k_\mathrm{r_{max}W}$   & maximum growth rate in response to feed moisture concentration & [\si{\gram\per\day}]\\
		$k_\mathrm{W_{med}C1}$ & lowest feed moisture below which the growth ceases & [\si{\gram\per\gram}]   \\
		$k_\mathrm{W_{med}C2}$ & feed moisture above which the ingestion rate can reach maximum & [\si{\gram\per\gram}]   \\
		$k_\mathrm{W_{med}C3}$ & feed moisture above which the diffusion of oxygen/air exchange starts to cease & [\si{\gram\per\gram}]   \\
		$k_\mathrm{W_{med}crit}$ & feed moisture  above which the larvae begins to die & [\si{\gram\per\gram}]   \\
		$k_\mathrm{r_{max}A}$  	& maximum observed growth rate in response to airflow rate & [\si{\gram\per\second}]  \\
		$k_\mathrm{A_{half}}$  	& air flow rate for which the growth rate is reduced to half & [\si{\litre\per\minute}]  \\
	\end{longtable}
}

\subsubsection{Larvae growth and dry mass partitioning}
The collective processes that define the growth and development of an organism are known to be the metabolism. 
The important processes of the metabolism in an organism ---here abstracted---are assimilation, maintenance, growth, maturity.
Using the mass/energy balance approach, these abstracted metabolic processes can be used to describe the growth and development rate of an organism.
General consideration in this approach is that the afore mentioned abstract processes can either use energy or mass for developing the model.
Since mass is easy to measure in-situ both in laboratory and in production compared to measuring energy, in this work models are derived based on the mass.

\begin{figure}[!h]
	\includegraphics[width=\linewidth,trim={25 0 0 0},clip]{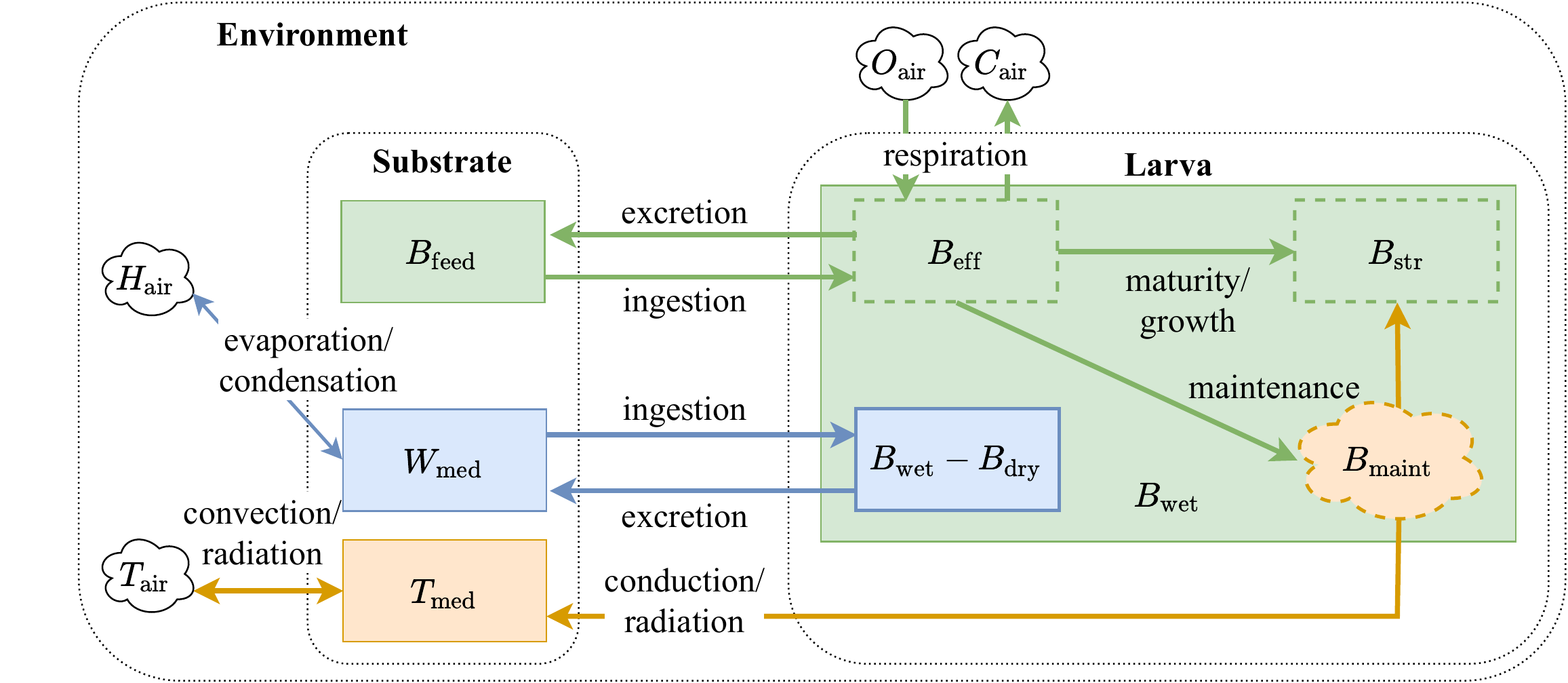}
	\caption{{\bf Mass and energy flow between the larva, substrate and the growing environment.}
	The rectangles represent the different states and the arrows indicate the flow of mass and energy (fluxes) between these states.
		Biomass and water in the substrate enters and exits larvae by ingestion and excretion.
		Gas exchange as a result of metabolic respiration takes place between the larva and the environment.
		Assimilated biomass and reserves $B_\mathrm{eff}$ is further converted into structure towards the larval maturity $B_\mathrm{str}$ and energy $B_\mathrm{maint}$ necessary for maintenance of the structure.
		The states represented in dashed lines indicate that they are not directly measurable unlike the larva wet and dry mass $B_\mathrm{wet}$ and $B_\mathrm{dry}$ respectively.
		A part of the $B_\mathrm{maint}$ is converted to heat, a byproduct of metabolism, and is lost to the substrate increasing its temperature $T_\mathrm{med}$.}
	\label{fig:larvae_mass_flow}
\end{figure}

Growth, which describes the increase in structural volume or mass in an organism, requires ingestion of feed.
This feed ingested by larva is converted into assimilates through the process of assimilation consuming a portion of the assimilates for this process. 
The assimilates are converted into structural mass towards growth and maturity.
Maturity, which is an indicator for larval development, consumes assimilates throughout the larval stage.
Maintenance respiration that keeps the organism alive also consumes some of the assimilates.
Using Forrester diagram, the flow and paritioning of biomass and energy through the afore mentioned processes are presented in Fig~\ref{fig:larvae_mass_flow}.

With this context and background, growth or the rate change of the larval dry mass is represented using mass balance model as
\begin{equation}\label{eq:dB_dry/dt}
\frac{\mathrm{d} B_\mathrm{dry}}{\mathrm{d}t}= \rlap{$\underbrace{\phantom{\phi_{\mathrm{B_{ing}}} - \phi_{\mathrm{B_{excr}}}  - \phi_{\mathrm{B_{assim}}}}}_{\phi_{\mathrm{B_{eff}}}}$} \phi_{\mathrm{B_{ing}}} - \phi_{\mathrm{B_{excr}}}  -
\overbrace{\phi_{\mathrm{B_{assim}}} - \phi_{\mathrm{B_{mat}}} - \phi_{\mathrm{B_{maint}}}}^{\phi_{\mathrm{B_{metab}}}},
\end{equation}
where $\phi_{\mathrm{B_{ing}}}$ is feed flux from substrate into the larvae, $\phi_{\mathrm{B_{excr}}}$ is the flux of non digested feed back to the substrate, $\phi_{\mathrm{B_{assim}}}$ is the feed converted into energy necessary for assimilation of the ingested feed, $\phi_{\mathrm{B_{maint}}}$ is the assimilates converted into energy for basal maintenance of existing structure, $\phi_{\mathrm{B_{mat}}}$ is the assimilates spent for growth and maturity (responsible for accumulating new structure) and $\phi_{\mathrm{B_{metab}}}$ represents all the assimilates consumed for metabolic activities.

The effective assimilates available for growth and maintenance $\phi_{\mathrm{B_{eff}}}$ can be expressed as
\begin{align}\label{eq:epsilon_assim}
\phi_{\mathrm{B_{eff}}} = \underbrace{(1-k_\mathrm{\alpha_{excr}} - k_\mathrm{\alpha_{assim}})}_{\epsilon_\mathrm{inges}} \ \phi_{\mathrm{B_{ing}}},
\end{align}
where $k_\mathrm{\alpha_{excr}}$ and $k_\mathrm{\alpha_{assim}}$ are respectively the fractions of feed excreted and spent in the process respectively and $\epsilon_\mathrm{inges}$ corresponds to the efficiency of the digested feed and provides information related to the quality of the feed.

Furthermore, maturity and maintenance expenses ($\phi_{\mathrm{B_{mat}}}$ and $\phi_{\mathrm{B_{maint}}}$) cannot be distinguished since these two processes are active during the entire development phase of the larvae \cite{Gligorescu2018}. Due to this reason as well as the terms being difficult to separately measure, they are combined into one flux term.
These flux components corresponding to the assimilation and maintenance are considered proportional to the weight/size of the organism \cite{Bertalanffy1957}. Therefore, Eq~\eqref{eq:dB_dry/dt} can be rewritten in terms of dry weight
as
\begin{equation}\label{eq:dB_dry/dt_ext}
\frac{\mathrm{d}B_\mathrm{dry}}{\mathrm{d}t}= \epsilon_\mathrm{inges} \ k_\mathrm{inges} \ B_\mathrm{dry} - \ k_\mathrm{maint}\ B_\mathrm{dry},
\end{equation}
where $k_\mathrm{inges}$ and $k_\mathrm{maint}$ are the specific assimilation rate and specific growth and maturity maintenance respectively (\si{\gram} feed \si{\gram}$^{-1}$ larvae \si{\second}$^{-1}$ in dry matter). The Eq~\eqref{eq:dB_dry/dt_ext} is of the form similar to the general form of von Bertalanffy model given in Eq~(5) of \cite{Bertalanffy1957} with $m=1$ for insects as suggested in \cite{Shi2014}. This model given by Eq~\eqref{eq:dB_dry/dt_ext} is rudimentary, describing only partitioning of the biomass across different biological processes.
To achieve the model goals described previously, it is also necessary to model various factors that regulate the rate of flow of the mass and energy fluxes across different processes.
Modelling of these factors are presented in the following sections.

\subsubsection{Larvae development}
Growth process of the larvae takes place in stages which is commonly known as instars with the \textit{Hermetia illucens} larvae undergoing a total of 6-7 instars \cite{Schremmer1984,Kim2010}.
The larvae undergo these development changes when the conditions necessary for growth are suitable.
This development, however, seems to be not dependent on the size of the larvae.
This can be inferred from the data presented in \cite{Diener2009}, where the larvae completed their development despite not reaching nearly half their maximum size.
Therefore, an alternate mechanism is required to model the developmental stages of the larvae.

The use of temperature sums or degree days to track the developmental stages for tracking the growth of an organism, including plants and insects, can be seen in literature \cite{Akers1984}.
In such applications, temperature sums served as an indicator to the total energy that organisms received in their lifetime.
However, this might only work for cases where the resources such as food, air concentration and heat are not limited.
Therefore, not only temperature but also other environmental conditions such as feed density, air concentration etc., needs to be considered to obtain virtual reference to the total energy received.
In this work, such indicator for total energy is obtained by using the sum of suitable growing conditions that the larvae receives for its development in its lifetime.
This sum, referred to as development sums in this work, can therefore be written as a function of all factors that affect the development as
\begin{equation}\label{eq:dT_sigma/dt_ext}
\frac{\mathrm{d}T_\mathrm{\Sigma}}{\mathrm{d}t}= \ rc_\mathrm{F}(B_\mathrm{feed}) \ rc_\mathrm{W}(W_\mathrm{med})  \ rc_\mathrm{A}(O_\mathrm{air}) \ rc_\mathrm{T}(T_\mathrm{med})  k_\mathrm{dev_{ts}}
\end{equation}
where $rc_\mathrm{F}$, $rc_\mathrm{W}$, $rc_\mathrm{A}$ and $rc_\mathrm{T}$ are the rate constants that describe the influence of feed, water, air concentration, and temperature on the development rates respectively and $k_\mathrm{dev_{ts}}$ constant to convert the development rate in development hours.
This Eq~\eqref{eq:dT_sigma/dt_ext} in effect provides a new dimension similar to age but calculated based on the rate constants of influencing factors. Modeling of these influences are presented in the next section but first the changes to the feeding and maturation process in response to the development sums are presented.

\paragraph{Feeding and growth stage}
Larval structural growth is a result of constant assimilation---the main function of the larvae is to accumulate enough assimilates and mass---lasting up-to the 6th instar \cite{Schremmer1984}.
According to the results published in \cite{Gligorescu2019}, the larvae assimilates the feed at highest rates during the 1st to 4th instar.
This gradually drops from 4th to 6th instar and assimilation finally ceases before 7th instar.
It is also observed in the works of \cite{Gligorescu2019}, that the mouth parts of the larvae undergo morphological changes suggesting changes in feeding behavior.

With respect to the model considered in this work, this transition into non-feeding stage indicate that the assimilation is maximum in the early larval stages, gradually decreases with increase in mass, and finally ceasing when the feeding stage is completed.
This transition of the assimilation process over the development stages, indicated by $T_\Sigma$, can be described as
\begin{equation}\label{eq:r_assim}
	r_\mathrm{B_{assim}}(T_\Sigma) = \begin{cases}
		r_\mathrm{assim_{max}}(B_\mathrm{dry})	& \text{if }  T_\Sigma < k_\mathrm{T_\Sigma 1} \\
   r_\mathrm{assim_{max}}(B_\mathrm{dry}) \left(\dfrac{T_\Sigma - k_\mathrm{T_\Sigma 2}}{k_\mathrm{T_\Sigma 1} - k_\mathrm{T_\Sigma 2}}\right) 		& \text{if }  k_\mathrm{T_\Sigma 1} \leq T_\Sigma < k_\mathrm{T_\Sigma 2} \\
   0 		& \text{if } T_\Sigma > k_\mathrm{T_\Sigma 2} \\
\end{cases},
\end{equation}
with $r_\mathrm{assim_{max}}(B_\mathrm{dry}) = 1-\dfrac{B_\mathrm{dry}}{k_\mathrm{B_{asy}}}$, where $k_\mathrm{B_{asy}}$ is the maximum asymptotic mass of the larvae, $k_\mathrm{T_\Sigma 1}$ is the transition point until which the larvae feeds at a maximum rate and $k_\mathrm{T_\Sigma 2}$ is the point beyond which the feeding comes to a halt. The function $r_\mathrm{assim_{max}}$ represents the ingestion/feeding potential of the larvae in relation to it its current size and the maximum size it can reach when infinitely fed.
With this Eq~\eqref{eq:r_assim}, a relation between the development sums and the size dependent ingestion is established.
 
\paragraph{Maturation stage}
In the final larval instar stage, the accumulated assimilates and reserves (e.g. fats) are further spent in developing the parts necessary to reach the maturity and transform into a pupae.
This maturity process was studied in \cite{Liu2017} which indicates a drop in the dry mass and, in specific, the crude fats during the transition from prepupae to pupae.
However, maturation ceases at the end of this transformation and then the pupal stage begins.
This maturity allocation, can be modeled as a rate that allocates the assimilates and reserves to the maturation process.
Allocation to maturity can be also seen in the modeling approaches of DEB \cite{Kooijman2010} (see Section 2.4).
In this work, such scheduling of assimilates to maturation is done such that the maturity process continues further after the feeding phase and until the larvae turns into a pupae. This continued allocation describes the drop in mass and indicates the change in body composition.
Therefore, the maturity allocation is described as
\begin{equation}\label{eq:r_mat}
r_\mathrm{B_{mat}}(T_\Sigma) = \begin{cases}
1 		& \text{if }  T_\Sigma < k_\mathrm{T_\Sigma 3} \\
0 		& \text{if } T_\Sigma \geq k_\mathrm{T_\Sigma 3} \\
\end{cases},
\end{equation}
where $k_\mathrm{T_\Sigma 3}$ indicates the end of prepupal or beginning of pupal stage.
For the model to track pupal development, the rate of maturity allocation shall be replaced with a non zero value since the pupae undergoes further metamorphosis consuming reserves.

\subsubsection{Effect of external factors on larval growth and development}
Factors that affect the growth and development of the larvae considered in this work include temperature, feed density, feed quality, moisture and air concentration.
Each of these parameters influence the larval development through various biological processes.
An attempt is made to model these influences through mechanistic and analytical models both from literature and developed based on the analysis of aggregated literature and experiment data.

\paragraph{Temperature}
Temperature has a direct effect on all the biochemical reactions that take place in the larvae and thus affecting its growth and development rate.
The effect of temperature on the growth of larvae can be modeled using Arrhenius equation \cite{Glasstone1941}.
Corrections to the metabolic rates for the temperatures beyond the upper and lower boundaries can be applied as
\begin{align}\label{eq:arrhenius}
r_\mathrm{T}(T_\mathrm{med}) = \dfrac{k_\mathrm{r_{ref}T} \exp\left( \dfrac{k_\mathrm{T_A}}{k_\mathrm{T_{ref}}} - \dfrac{k_\mathrm{T_A}}{T_\mathrm{K}} \right)}
{\left(1 + \exp\left(\dfrac{k_\mathrm{T_{AL}}}{T_\mathrm{K}} - \dfrac{k_\mathrm{T_{AL}}}{k_\mathrm{T_L}}\right)
	+ \exp\left(\dfrac{k_\mathrm{T_{AH}}}{k_\mathrm{T_H}} - \dfrac{k_\mathrm{T_{AH}}}{T_\mathrm{K}}\right)\right)}
\end{align}
with $T_\mathrm{K} = T_\mathrm{med}$ in \si{\kelvin}, where $k_\mathrm{T_A}$, $k_\mathrm{T_{AL}}$ and $k_\mathrm{T_{AH}}$ are Arrhenius temperatures at the reference  $k_\mathrm{T_{ref}}$,  lower boundary $k_\mathrm{T_L}$ and upper boundary $k_\mathrm{T_H}$ temperatures respectively and $k_\mathrm{r_{ref}T}$ is the known reference rate.

Another analytical model proposed in \cite{Logan1976} (see Eq~(10) of \cite{Logan1976}), referred to in this work as Logan-10, also describes the growth rate in response to the temperature as well considering the effects of denaturation and desiccation at high temperatures.
The Logan-10 model was modified such that for temperatures beyond the upper threshold, the resulting growth is zero instead of a negative growth. The resulting modified Logan-10 model is given as
\begin{equation}\label{eq:logan-10mod}
r_\mathrm{T}(T_\mathrm{med}) = k_\mathrm{r_{max}T} \left(1 + k_\mathrm{\gamma}\exp\left(- k_\mathrm{\rho T} \left(T_\mathrm{med} - k_\mathrm{T_{base}}\right)\right) + \exp\left(-\dfrac{k_\mathrm{T_{max}} - T_\mathrm{med}}{k_\mathrm{\Delta T}}\right)\right)^{-1},
\end{equation}
with $k_\mathrm{\gamma} = \left(\frac{k_\mathrm{r_{max}T} - k_\mathrm{r_{base}T}}{k_\mathrm{r_{base}T}}\right)$, where $k_\mathrm{r_{max}T}$ is the maximum observed rate (\si{\per\second}), $k_\mathrm{r_{base}T}$ is the minimum rate at the temperature above the lower threshold, $k_\mathrm{\rho T}$ rate change in response to the temperature, $k_\mathrm{T_{max}}$ is the lethal maximum temperature and $k_\mathrm{\Delta T}$ is the width of the high temperature boundary layer.
The Logan-10 model has one less parameter compared to Arrhenius model presented in Eq~\eqref{eq:arrhenius} and can be intuitively approximated from the available data. In this work, both these models will be evaluated and the corresponding parameters will be estimated.

\paragraph{Feed Density}
The feed flux assimilated by the larvae is effected by few factors, considered important in this work due to its application, such as feed availability and change in feeding behavior due to the modifications to the mouth parts of the larvae in its final instars.
It is common in literature to model the change in ingestion rate due to substrate availability using a type II function.
Monod presented an adaptation of this function to model the growth of bacterial cultures in \cite{Monod1949}.
This Monod equation is adapted in this work as
\begin{equation}\label{eq:feed_monod}
r_\mathrm{F}(B_\mathrm{feed}) =  k_\mathrm{r_{max}dm} \dfrac{B_\mathrm{feed}}{B_\mathrm{feed} + k_\mathrm{B_{half}dm}},
\end{equation}
where $B_\mathrm{feed}$ is the feed density in the substrate/growing medium, $k_\mathrm{r_{max}dm}$ is the maximum development rate (\si{\per\second}) at highest feed density and $k_\mathrm{B_{half}dm}$ is the feed density resulting in half of the maximum rate.

Similarly, Eq~\eqref{eq:feed_monod} can be rewritten to also model the growth rate of the larvae as
\begin{equation}\label{eq:feed_monod_growth}
r_\mathrm{F_{grw}}(B_\mathrm{feed}) =  k_\mathrm{r_{max}gm} \dfrac{B_\mathrm{feed}}{B_\mathrm{feed} + k_\mathrm{B_{half}gm}},
\end{equation}
where $k_\mathrm{r_{max}gm}$ is the maximum growth rate (\si{\gram\per\second}) of the larvae and $k_\mathrm{B_{half}gm}$ is the feed density resulting in half of the maximum growth rate.

\paragraph{Feed Moisture}
Based on the data presented in \cite{Palma2018}, the authors suggest that the moisture of the feed (\si{\kilogram} water in \si{\kilogram} wet feed) has a first order effect. However, in that study, data was only available for the moisture content of \SI{48}{}-\SI{68}{\percent}.
Another early work studied the development of different flies, including \textit{Hermetia illucens}, under different substrate moisture conditions \cite{Fatchurochim1989}.
In this work, the authors concluded that the development increased with increase in moisture content between \SI{30}{}-\SI{70}{\percent}, while, at 20, 80 and \SI{90}{\percent} there was no development observed.
The moisture experiment performed as part of this work, covered the feed moisture in the very low and high concentrations.
Based on these results, the feed moisture has different influences on the larval growth.
Firstly, with lower moisture, the feed may not be ingestible and thus result in slower growth and high mortality rate at very low moisture levels.
Secondly, with increasing moisture, feed could be assimilated better resulting in better growth.
Finally, at higher moisture concentrations, intake of oxygen might be reduced resulting in slower growth and higher larval mortality.

Based on these observations, the influence of water content in feed on the larval growth
can be modelled as
\begin{equation}\label{eq:r_w}
	r_\mathrm{W}(W_\mathrm{med\%}) = k_\mathrm{r_{max}W} \ r_\mathrm{W_{assim}}(W_\mathrm{med\%}) \ r_\mathrm{W_{resp}}(W_\mathrm{med\%}),
\end{equation}
	where $k_\mathrm{r_{max}W}$ is the maximum growth rate, and the influence of moisture content on the assimilation rate and respiration rate $r_\mathrm{W_{assim}}$ and $r_\mathrm{W_{resp}}$ respectively are modelled as
\begin{align}
r_\mathrm{W_{assim}}(W_\mathrm{med\%}) &= \begin{cases}
	\quad 0 		& \text{if }  W_\mathrm{med\%} \leq k_\mathrm{W_{med}C1} \\
\dfrac{W_\mathrm{med\%} - k_\mathrm{W_{med}C1}}{k_\mathrm{W_{med}C2} - k_\mathrm{W_{med}C1}} & \text{if } k_\mathrm{W_{med}C1} < W_\mathrm{med\%} < k_\mathrm{W_{med}C2} \\
\quad 1		& \text{if }  W_\mathrm{med\%} \geq k_\mathrm{W_{med}C2}
\end{cases},\label{eq:r_w_assim} \\
r_\mathrm{W_{resp}}(W_\mathrm{med\%}) &= \begin{cases}
	\quad 1		& \text{if }  W_\mathrm{med\%} \leq k_\mathrm{W_{med}C3} \\
	\dfrac{W_\mathrm{med\%} - k_\mathrm{W_{med}crit}}{k_\mathrm{W_{med}C3} - k_\mathrm{W_{med}crit}} & \text{if } k_\mathrm{W_{med}C3} < W_\mathrm{med\%} < k_\mathrm{W_{med}crit} \\
	\quad 0 		& \text{if }  W_\mathrm{med\%} \geq k_\mathrm{W_{med}crit}
\end{cases},\label{eq:r_w_resp}
\end{align}
where $k_\mathrm{W_{med}C1}$ is the lowest water content in the feed below which growth ceases due to reduced feed ingestion rate, $k_\mathrm{W_{med}C2}$ is the moisture concentration above which the ingestion rate is maximum, $k_\mathrm{W_{med}C3}$ is the water concentration above which diffusion of air into substrate and thus the larvae drops, and $k_\mathrm{W_{med}crit}$ is the highest water concentration above which oxygen diffusion ceases.
 
\paragraph{Air concentration}
Effect of aeration in the growing environment influences the larval growth through the availability of O$_2$ necessary for respiration.
A study performed in \cite{Palma2018} that compared the larval growth at different air flow rates and thus the available O$_2$ concentration used a logistic model to describe the data.
\begin{equation}\label{eq:r_O_log}
r_\mathrm{A}(A) = k_\mathrm{r_{max}A}\left( 1 + \exp \left(\dfrac{A - k_\mathrm{A_{inf}}}{k_\mathrm{A_{trans}}}\right)\right)^{-1},
\end{equation}
where $k_\mathrm{r_{max}A}$ is the maximum rate, $k_\mathrm{A_{inf}}$ is the infliction point and $k_\mathrm{A_{trans}}$ is the slope.
However, in \cite{Richard2006} the authors despite highlighting the logistic model, have used a type II function to model the development rate based on the O$_2$ concentration.
In this work, the influence of O$_2$ concentration  is considered as a resource necessary for the underlying biological processes and therefore modeled as  
\begin{equation}\label{eq:r_O_monod}
r_\mathrm{A}(A) = k_\mathrm{r_{max}A} \frac{A}{A + k_\mathrm{A_{half}}},
\end{equation}

where $k_\mathrm{r_{max}A}$ is the maximum growth rate under certain high O$_2$ concentrations (Air flow rate) and $k_\mathrm{A_{half}}$ is the airflow rate for which the growth rate is reduced by half.

 \subsubsection{Combining growth, development and external factors}
\begin{figure}[!h]
	\includegraphics[width=\linewidth,trim={25 0 0 0},clip]{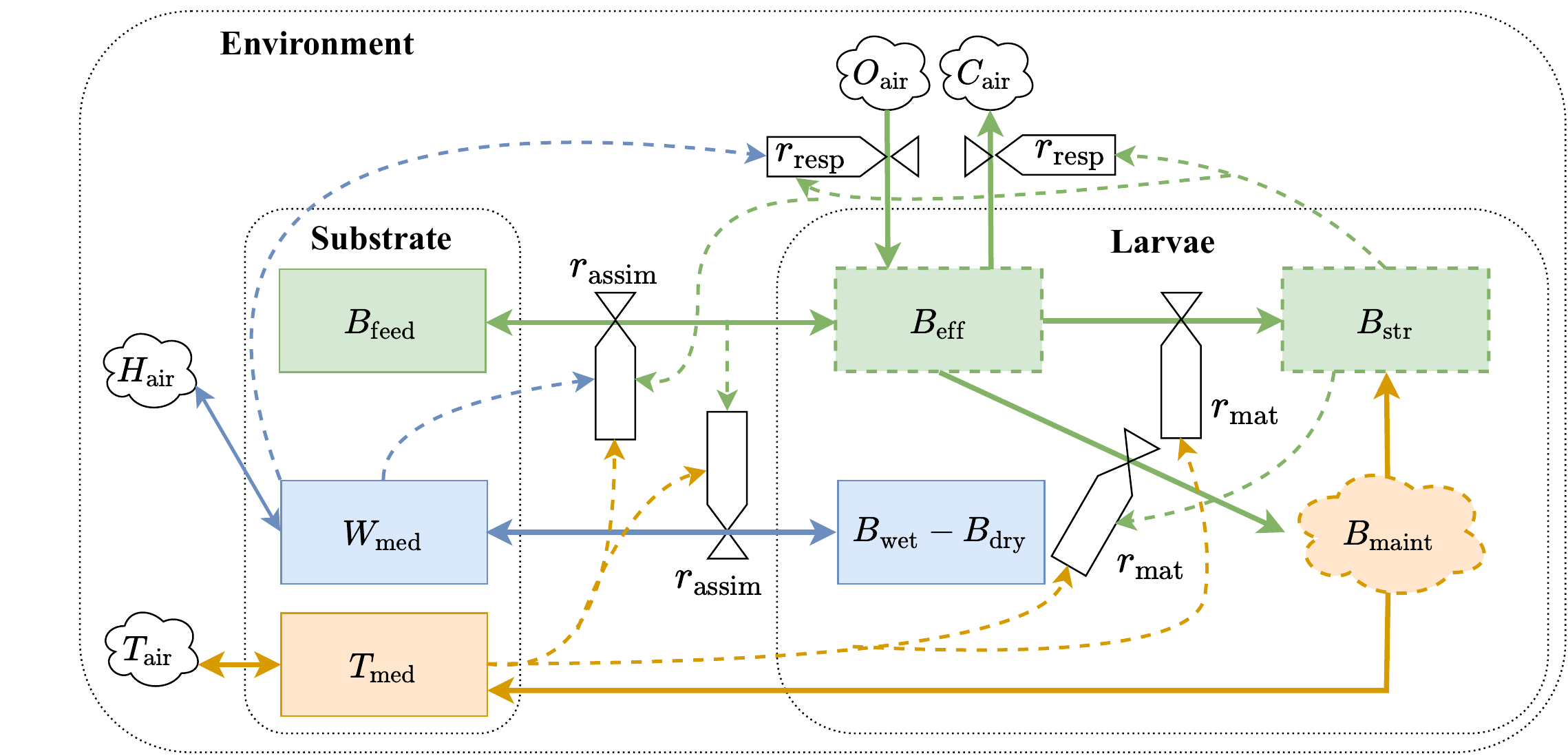}
	\caption{{\bf Mass and energy transfer.}
	The flow of mass and energy between the substrate or growing medium, larva body and the environment in response to various states and environment conditions are represented using the valves that regulate this flow.
	Influence of the states on the rate are indicated using dashed lines.
	Influence of the states on the rates are not explicitly indicated when the flow takes place between those corresponding states.
	Rates $r_\mathrm{assim}$, $r_\mathrm{mat}$, and $r_\mathrm{resp}$, represent the assimilation, maturity-maintenance and respiration as a function of various states that influence the flow of biomass and energy.}
	\label{fig:larvae_substrate_dynamics}
\end{figure}

The considered factors affecting the growth of larvae and the movement of mass and energy between larva, growing medium and the environment is summarized in Fig~\ref{fig:larvae_substrate_dynamics} using Forrester diagram.
The $B_\mathrm{str}$, representing the structural mass of the larva, has its influences on most of the rate flows as seen in Fig~\ref{fig:larvae_substrate_dynamics}.

With the knowledge of factors affecting the different biological processes, here, we apply the obtained rate functions Eq~\cref{eq:r_assim,eq:r_mat,eq:logan-10mod,eq:feed_monod_growth,eq:r_O_monod,eq:r_w} to the corresponding mass and energy flow.
The combined rates and states represent the larvae growth as
 \begin{equation}\label{eq:dB_dry_rate}
 \frac{\mathrm{d}B_\mathrm{dry}}{\mathrm{d}t}= \underbrace{\epsilon_\mathrm{inges} \ rc_\mathrm{F_{grw}} \ rc_\mathrm{W} \  rc_\mathrm{A} \ rc_\mathrm{T}\ r_\mathrm{B_{assim}} \ k_\mathrm{inges} \ B_\mathrm{dry}}_{\text{effective assimilation}}  - \ \underbrace{rc_\mathrm{F_{grw}} \ rc_\mathrm{A} \ rc_\mathrm{T} \ r_\mathrm{B_{mat}} \  k_\mathrm{maint}\ B_\mathrm{dry}}_{\text{maturity-maintenance}}.
 \end{equation}
 Further, rearranging the terms of the above equation, for simplification, results in
 \begin{equation}\label{eq:dB_dry_final}
 \frac{\mathrm{d}B_\mathrm{dry}}{\mathrm{d}t}= \left(r_\mathrm{assim} \ \epsilon_\mathrm{inges} \ k_\mathrm{inges} - r_\mathrm{mat} \ k_\mathrm{maint}\right) B_\mathrm{dry},
 \end{equation}
 with $r_\mathrm{assim} = r_\mathrm{B_{assim}} \ rc_\mathrm{F_{grw}}  \ rc_\mathrm{W}  \ rc_\mathrm{A} \ rc_\mathrm{T}$ models the factors affecting the assimilation process rate and  $r_\mathrm{mat} = r_\mathrm{B_{mat}} \ rc_\mathrm{T} \ rc_\mathrm{A} \ rc_\mathrm{F}$ models the factors affecting the maturity and maintenance.
 The rate constants $rc_\mathrm{F}$, $rc_\mathrm{F_{grw}}$, $rc_\mathrm{W}$, $rc_\mathrm{A}$ and $rc_\mathrm{T}$ are obtained by normalizing the rate functions $r_\mathrm{F}$, $r_\mathrm{F_{grw}}$ $r_\mathrm{W}$, $r_\mathrm{A}$, and $r_\mathrm{T}$  presented in Eq~\cref{eq:logan-10mod,eq:feed_monod,eq:feed_monod_growth,eq:r_O_monod,eq:r_w} respectively.

\paragraph{Implementation of switching functions}
All functions modelled in this work as cases or switching functions, Eq~\eqref{eq:r_assim}, \eqref{eq:r_mat}, \eqref{eq:r_w_assim}, and \eqref{eq:r_w_resp}, are realized as logistic functions for a smooth transitioning between the conditions.
The parameters defining the boundary conditions are replaced in the logistic function, for example, for Eq~\eqref{eq:r_assim} as 
\begin{equation}\label{eq:metab_switch}
	r_\mathrm{B_{assim}}(T_\Sigma) = \left( 1 + \exp \left(-4\left(\dfrac{T_\Sigma - k_\mathrm{T_\Sigma inf}}{k_\mathrm{T_\Sigma 1} - k_\mathrm{T_\Sigma 2}}\right)\right)\right)^{-1}
	\end{equation}
	with $k_\mathrm{T_\Sigma inf} = k_\mathrm{T_\Sigma 1} + 0.5(k_\mathrm{T_\Sigma 2} - k_\mathrm{T_\Sigma 1})$.

\subsection{Model validation and parameter estimation}
Models presented in this work are mostly nonlinear and therefore, nonlinear least squares data fitting method was used for parameter estimation.
This was formulated as an optimization problem with the objective of finding the parameter that minimizes the sum of square of errors as
\begin{equation*}\label{eq:lsqnonlin}
\min_{\mathbf{p}}\, \sum_{i} (f(\mathbf{p},\mathbf{X}_i) - \mathbf{Y}_i)^2,
\end{equation*}
where $\mathbf{p}$ represents the parameters to be estimated, $f(\mathbf{p},\mathbf{X})$ represents the model, $\mathbf{Y}$ is the measured data and $i$ represents the measurement samples.
This parameter estimation problem was implemented in MATLAB using the $lsqcurvefit$ function in a multi-search framework to explore possible solutions within the specified boundary values of parameters.
Simulation of dynamic models were performed using the $ode45$ solver in MATLAB to solve the differential equations.

The data sets obtained from different literature sources, as presented in Table~\ref{tab:Datasource}, were used in this work for the purpose of parameter estimation and validation.
Data sets originating from different sources but used for common model and parameter estimation were first normalized between 0 and 1 on the Y-axis to form data collection.
Such data collection were used for the parameter estimation process of each model.
Also, the parameter values presented in this work are obtained using the average values of the normalized data collection.
Parameter estimation and the resulting quality of fit for each data set is presented in the following section.
\begin{table}[!ht]
	\begin{adjustwidth}{-2.25in}{0in} 
		\centering
		\caption{
			{\bf Data sets and their source used for model validation and parameter estimation}}
		\begin{tabular}{|c|m{0.15\textwidth}|m{0.6\textwidth}|m{0.4\textwidth}|}
			\hline
			\bf Dataset ID & \bf Source & \bf Description & \bf Application\\
			\thickhline
			T1, T2		& Fig~3 of \cite{Chia2018} 		& \multirow{2}{=}{Development time of BSF on different diets at different temperatures} 	& \multirow{2}{=}{Validation and parameter estimation for Eq~\cref{eq:arrhenius,eq:logan-10mod}}\\ \cline{1-2}
			T3, T4		& Table~1 of \cite{Shumo2019} 		&  & \\ \hline
			F1			& Table~2, 3 of \cite{Diener2009} 	& Development time and dry weight respectively of BSF larvae under different feeding rates & \multirow{2}{=}{Validation and parameter estimation for Eq~\cref{eq:feed_monod,eq:feed_monod_growth}}\\ \cline{1-3}
			F2-F4	& Table~2, Fig~1 of \cite{Barragan-Fonseca2018} 	& Development time and dry weight respectively of BSF larvae under different feed and feeding rates& \\ \hline
			M1			& Fig~4 of \cite{Palma2018} 	& \multirow{3}{0.6\textwidth}{ Larvae growth/dry weight change under different substrate moisture content} & \multirow{3}{=}{Validation and parameter estimation for Eq~\cref{eq:r_w,eq:r_w_assim,eq:r_w_resp}}\\ \cline{1-2}     
			M2		& This work && \\\cline{1-2}
		    M3		&Table~2 \cite{Fatchurochim1989} && \\\hline
		    A1			& Fig~2 of \cite{Palma2018} 	& Larvae growth/dry weight change under different aeration rate & Validation and parameter estimation for Eq~\cref{eq:r_O_monod,eq:r_O_log}\\\hline
		    G1			& Fig~1 of \cite{Liu2017} 	& Larvae growth/dry weight change over the developmental phases &  \multirow{3}{=}{Validation and parameter estimation for Eq~\cref{eq:dB_dry_final,eq:r_assim,eq:r_mat}}\\\cline{1-3}
		    D1, D5			& Fig~2 of \cite{Diener2009} & Larvae growth/dry weight change over the developmental phases under different feeding rates& \\\hline
		    D2-D4			& Fig~2 of \cite{Diener2009} & Larvae growth/dry weight change over the developmental phases under different feeding rates& Validation of Eq~\cref{eq:dB_dry_final}\\	    
			\thickhline                          
		\end{tabular}
		\label{tab:Datasource}
	\end{adjustwidth}
\end{table}

\section{Results and discussion}
In this section, firstly, the performance of the individual rate functions Eq~\cref{eq:arrhenius,eq:logan-10mod,eq:feed_monod,eq:feed_monod_growth,eq:r_O_log,eq:r_O_monod,eq:r_w} describing the influence of external factors on growth and development are presented.
Secondly, performance of the combined dynamic model representing the growth Eq~\eqref{eq:dB_dry_final} and development Eq~\eqref{eq:dT_sigma/dt_ext} are presented, highlighting the validity of the rate functions Eq~\cref{eq:r_assim,eq:r_mat} for assimilation and maturation respectively.
Finally, the dynamic growth and development model are validated using additional datasets.

\subsection{Temperature influence}
\begin{figure}[!h]
	\centering
	\includegraphics[width=\linewidth,trim={0 0 0 0},clip]{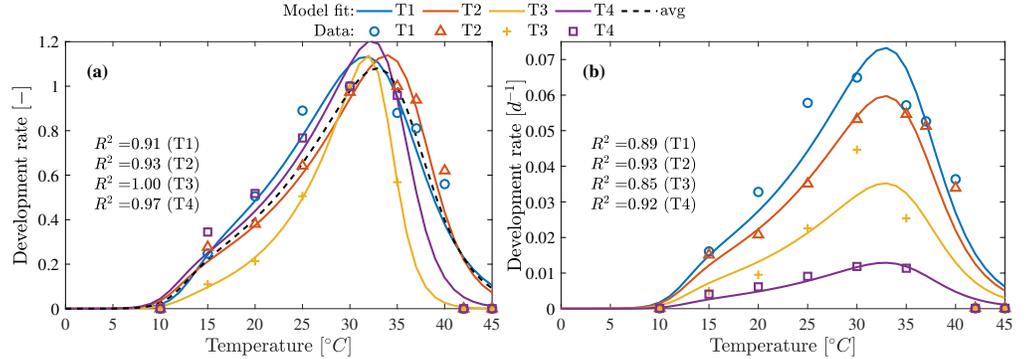}
	\caption{{\bf Temperature influence on development rate using Arrhenius model.} \textbf{(a)} Parameters estimation using Arrhenius model \eqref{eq:arrhenius} and normalized data. \textbf{(b)} Development rate estimation using the parameters estimated for data set obtained by averaging all (T1-T4) data sets. Model fit represented as avg shows the performance of the final model.}
	\label{fig:T_dependency_arrhenius}
\end{figure}

\begin{figure}[!h]
	\centering
	\includegraphics[width=\linewidth,trim={0 0 0 0},clip]{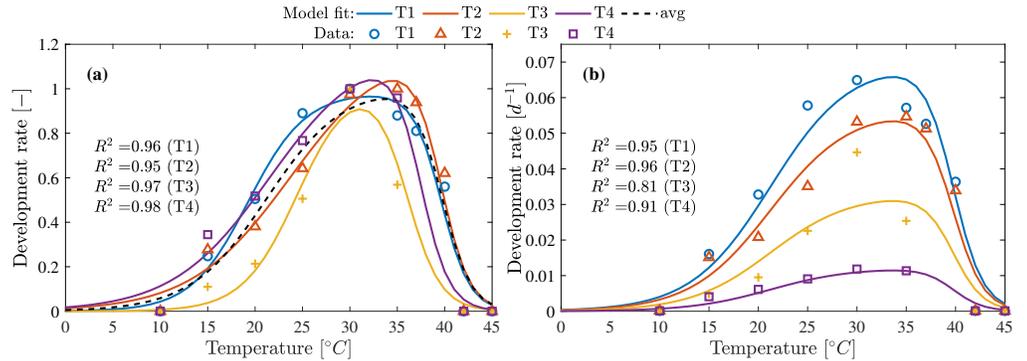}
	\caption{{\bf Temperature influence on development rate using modified Logan-10 model.} \textbf{(a)} Parameters estimation using modified Logan-10 model \eqref{eq:logan-10mod} and normalized data. \textbf{(b)} Development rate estimation using the parameters estimated for data set obtained by averaging all (T1-T4) data sets. Model fit represented as avg shows the performance of the final model.}
	\label{fig:T_dependency_logan}
\end{figure}

A total of four data sets (T1-T4) representing the influence of temperature on larvae development was obtained from \cite{Chia2018,Shumo2019}.
These four data sets represent the temperature dependency under four different feed types and was used to obtain the parameters for models Eq~\cref{eq:arrhenius,eq:logan-10mod}.
Fig~\ref{fig:T_dependency_arrhenius} shows the results of the parameter estimation using the Arrhenius model \eqref{eq:arrhenius} and Fig~\ref{fig:T_dependency_logan} for the modified Logan-10 model Eq~\eqref{eq:logan-10mod}. Both models perform well in describing the data with good quality of fit ($R^2>$0.91).
The model parameter obtained from the average data set was used to explain the data sets T1-T4 as shown in Fig~\ref{fig:T_dependency_arrhenius}(b) and Fig~\ref{fig:T_dependency_logan}(b).
The results of the model with the estimated parameters from the average data performed well in explaining all data sets with the only exception for T3 where both models could not explain the peak at \SI{30}{\celsius}.
Modified Logan-10 model has overall better quality of fit for both normalized and actual data sets. For temperatures below 15~\si{\celsius}, Eq~\ref{eq:arrhenius} provides a better fit at the expense of one additional parameter.

\subsection{Feed Density}
\begin{figure}[!h]
	\centering
	\includegraphics[width=\linewidth,trim={0 0 0 0},clip]{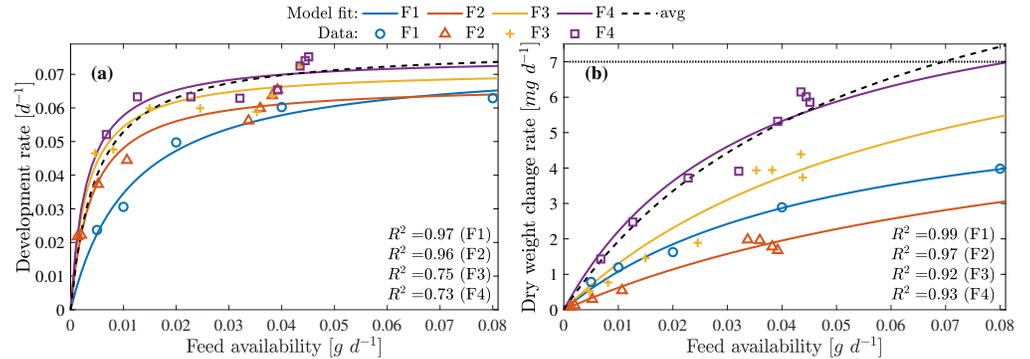}
	\caption{{\bf Feed availability on development and growth.} \textbf{(a)} Larvae development rates at varying feed availability. \textbf{(b)} Larvae growth rates at varying feed availability. Model fit avg represents the results of the average model scaled to the maximum observed development rate from the 4 data sets.}
	\label{fig:type2_feed}
\end{figure}

\begin{figure}[!h]
	\centering
	\includegraphics[width=\linewidth,trim={0 0 0 0},clip]{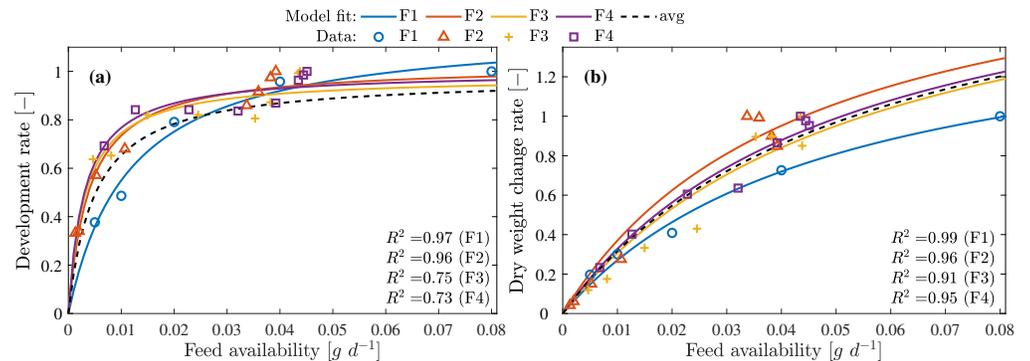}
	\caption{{\bf Feed availability on development and growth (Normalized).} \textbf{(a)} Larvae development rates at varying feed availability. \textbf{(b)} Larvae growth rates at varying feed availability. Mode fit avg represents the results of the average model for the combined data sets.}
	\label{fig:type2_feed_norm}
\end{figure}

Results published in \cite{Diener2009} was used to obtain data set F1 and \cite{Barragan-Fonseca2018} for data sets (F2-F4) representing the development and growth rate under different feeding densities and feed types.
The feed density defined in these works use gram dry mass of feed available/provided per larvae per day (\si{\gram\per\day} per larva) during the feeding periods.
These data sets were used to obtain the development rates and growth rates using the model Eq~\eqref{eq:feed_monod} and Eq~\eqref{eq:feed_monod_growth} as shown in Fig~\ref{fig:type2_feed} and Fig~\ref{fig:type2_feed_norm}.
The models describe accurately ($R^2=0.97$) for the data set F1 due to the availability of measurement for uniformly distributed feed densities.
In case of F2-F4, the data set also includes both batch fed and continuous fed experiment measurements resulting in scattered measurements and thus lower quality of fit.
Growth rate model, on the contrary, performs better in describing all data sets F1-F4 with $R^2>0.92$.
The parameters for these models are obtained by combining the normalized data sets F1-F4.
The resulting model from this combined data is shown with the dashed line (indicated as avg) as in Fig~\ref{fig:type2_feed_norm}.
From these results it could be concluded that these models can be used to compute the growth and development rates under different feeding rates.

\subsection{Moisture effect on growth}
\begin{figure}[!h]
	\centering
	\includegraphics[width=\linewidth,trim={0 0 0 0},clip]{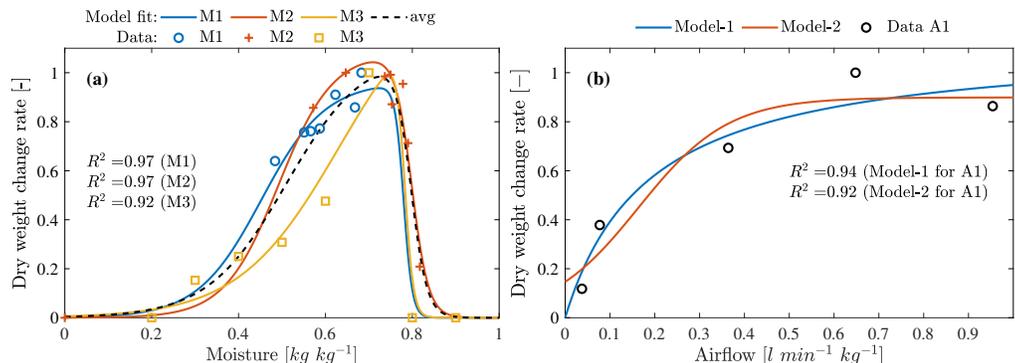}
	\caption{{\bf Moisture and airflow on growth rate.} \textbf{(a)} Effect of substrate/feed moisture on the growth rate. \textbf{(b)} Effect of airflow rate on the growth rate in closed production. Model-1 represents the Monod model Eq~\eqref{eq:r_O_monod} and Model-2 represents the logistic model Eq~\eqref{eq:r_O_log}.}
	\label{fig:type2_O2}
\end{figure}

To evaluate the model presented in Eq~\eqref{eq:r_w}, for influence of moisture on the growth, a total of three data sets M1-M3 were obtained.
M1 and M3 are results published in \cite{Palma2018} and \cite{Fatchurochim1989} respectively.
Data sets M1 and M2 does not contain measurements for the entire moisture concentration range but M3 provides data for the range from 20\% to 90\% as seen in Fig~\ref{fig:type2_O2}(a).
The proposed model is capable of describing the growth for the considered data sets and also the observations from the moisture experiment coincides with \cite{Fatchurochim1989} for the higher moisture concentration.
On the contrary, higher development rate was observed for moisture at 80\% in \cite{Cheng2017}.
Further investigation with complementary data sets may be necessary to identify the boundaries for higher moisture concentrations.

\subsection{Airflow rate}
Only one literature was found that included the effect of airflow on the growth of the larvae in \cite{Palma2018}.
In that work, closed 750~\si{\milli\liter} bioreactors with different aeration rates were used to study the larval growth rates.
As expected, in closed environment, growth was slow at lower aeration rates and increased gradually with increasing aeration rates and finally saturates. 
Model Eq~\eqref{eq:r_O_monod} and \eqref{eq:r_O_log} were evaluated and the results are presented in Fig~\ref{fig:type2_O2}(b).
From these results, one can see that both models can describe the growth under various aeration rates.
However, Eq~\eqref{eq:r_O_monod} provides better results for the available data sets.
Further studies might be necessary to obtain the growth response to different flow rates under varying moisture concentrations to identify correlation between them. 

\subsection{Larvae growth and development}

The larvae growth model presented in Eq~\eqref{eq:dB_dry_final}, describing the evolution of dry mass of the larvae, and the development model presented in Eq~\eqref{eq:dT_sigma/dt_ext} was validated based on the dry weight measurements presented in literature \cite{Diener2009} and \cite{Liu2017}.

\begin{figure}[!h]
	\centering
	\includegraphics[width=\linewidth,trim={0 0 0 0},clip]{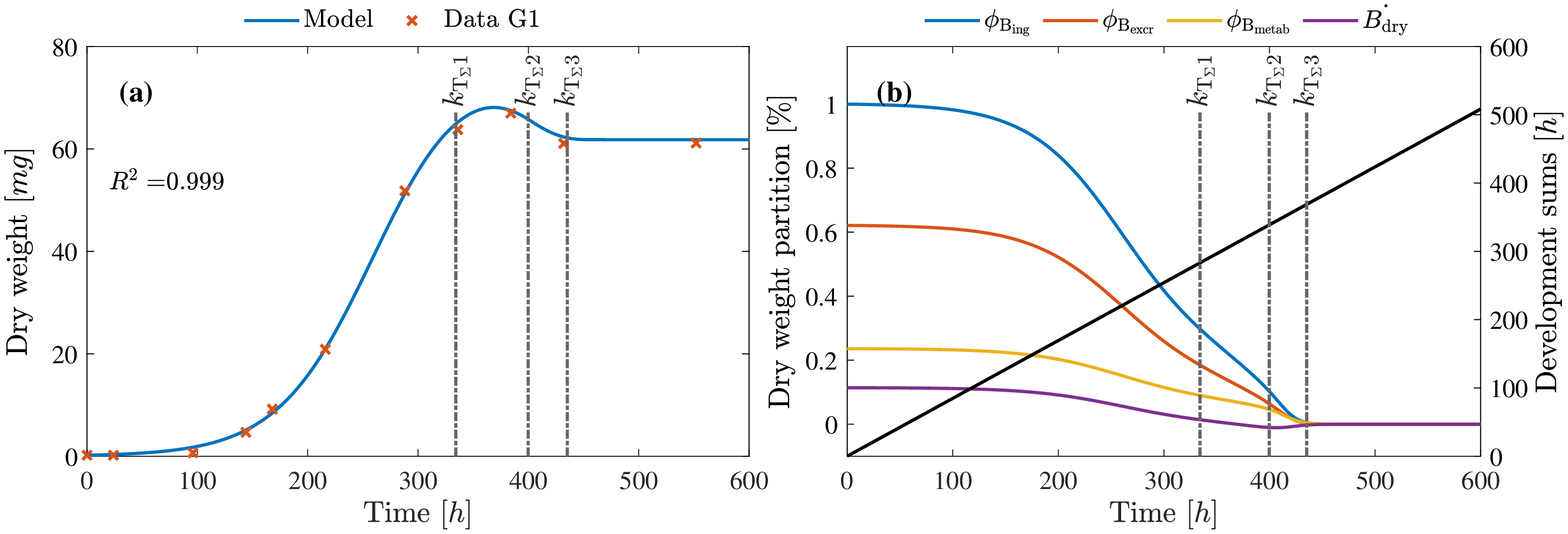}
	\caption{{\bf Larvae growth, development and biomass partitioning}. \textbf{(a)} Larvae dry mass evolution over time \textbf{(b)} Partitioning of assimilates and dry mass over development phases.
	The development sums where the assimilation and maturity transition, are indicated by the horizontal markers labelled $k_\mathrm{T_{\Sigma}1}$, $k_\mathrm{T_{\Sigma}2}$, and $k_\mathrm{T_{\Sigma}3}$.}
	\label{fig:biomass_balance}
\end{figure}

Dry mass of the \textit{Hermetia illucens} from eggs to adult, presented in \cite{Liu2017}, was used to obtain estimates of the parameters marking the important stages $k_\mathrm{T_\Sigma 1}$,$k_\mathrm{T_\Sigma 1}$, and $k_\mathrm{T_\Sigma 3}$.
Based on the biomass conversion efficiency for chicken feed provided in \cite{Diener2009, Bava2019}, parameters for dry mass distribution to metabolism, excretion and growth are also estimated to 0.24\%, 0.62\%, and 0.11\% respectively.
The performance of the model based on the estimated parameters is presented in Fig~\ref{fig:biomass_balance}.
Initially, the change of larval dry mass from the 1st larval instar to the 5th instar is regulated by the asymptotic size of the larva (in dry mass) as seen between 0-\SI{320}{\hour} marked by $k_\mathrm{T_\Sigma 1}$.
As the larvae approaches its last instar, ceasing of the ingestion process marked by the morphological changes such as modification of mouth parts and darkening of the skin are identified by the $k_\mathrm{T_\Sigma 2}$ and $k_\mathrm{T_\Sigma 3}$.
The final transition from prepupae to pupae is marked at the end of $k_\mathrm{T_\Sigma 3}$, indicating the end of larval growth and start of pupal stage.

\begin{figure}[!h]
	\centering
	\includegraphics[width=\linewidth,trim={0 0 0 0},clip]{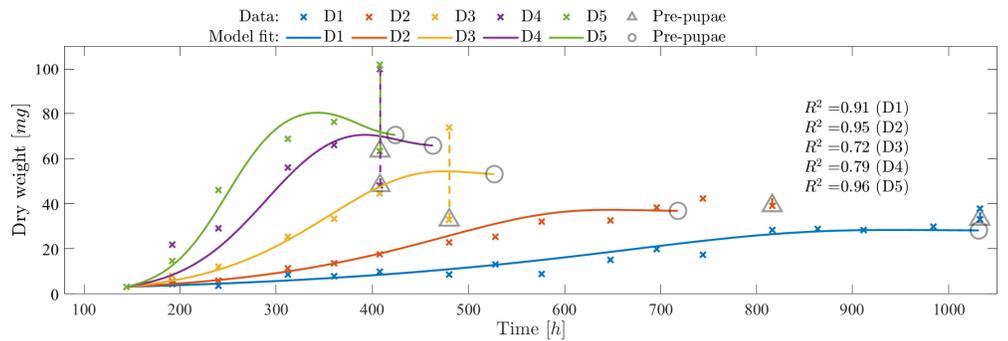}
	\caption{{\bf Validation of larvae growth and development model} Models are validated based on the data sets D1-D5 as published in \cite{Diener2009}. The vertical line indicates the time point where about 50\% of the larvae are transformed into prepupae. The circle on the corresponding model fit indicate the $k_\mathrm{T_\Sigma 3}$ time point when development of larvae are completed.}
	\label{fig:biomass_diener}
\end{figure}

Furthermore to validate the model for different data sets, data set from \cite{Diener2009} was used.
Using the data sets D1 and D5, model parameters were further adjusted for the new setup and using these new parameters the performance of the model was validated for all data sets D1-D5.
The results as seen in Fig~\ref{fig:biomass_diener}, highlights the performance of the model by providing the dry weight evolution as well as the indication of the different Larval development stages.

As seen from Fig~\ref{fig:biomass_diener}, the $R^2$ for the data sets D1,D2 and D5 are $>0.91$ but comparatively lower for D3 and D4.
This is purely due to the variance in the final recorded weight for D3 and D4.
To support this inference, we can also observe that for D3, D4 and D5 there were larvae with higher mass on the same day when 50\% prepupae were observed.
The Eq~\eqref{eq:dB_dry_final} also models higher assimilates allocation to maturity for higher feed density, indicating that more reserves are available in larvae growing in higher feed density.
This can be seen in the drop in dry mass when larvae transforms to prepupae.
This drop, as explained by the model and also as observed from the data is highest for D5 and lowest for D1.

\subsection{Summary}
From these observation one can conclude that the dynamic model developed in this work serves the two main intended goals: (1) model the larval growth through change of dry mass $B_\mathrm{dry}$ over its development stages under different growing conditions; and (2) model the transition of larval development through the development sums $T_\mathrm{\Sigma}$ under different growing conditions.
These results were achieved using simple model structures which plausibly and consistently explained various data from the literature.
Finally, the model parameters estimated in this work are summarized in Table~\ref{table:parameters}.

\begin{table}[!ht]
	\begin{adjustwidth}{-2.25in}{0in} 
		\centering
		\caption{
			{\bf Estimated model parameter values}}
		\begin{tabular}{|m{0.1\linewidth}m{0.18\linewidth}|m{0.1\linewidth}m{0.18\linewidth}|m{0.1\linewidth}m{0.18\linewidth}| }
			\hline
			\bf Parameter & \bf Est. value & \bf Parameter & \bf Est. value & \bf Parameter & \bf Est. value \\
			\thickhline
			${k_\mathrm{r_{ref}T}}^\text{\eqref{eq:arrhenius}}$& $\SI{0.7195}{}^*$ &
			${k_\mathrm{T_A}}^\text{\eqref{eq:arrhenius}}$& \SI{8450}{\kelvin} &
			${k_\mathrm{T_{AL}}}^\text{\eqref{eq:arrhenius}}$& \SI{60000}{\kelvin}\\
			${k_\mathrm{T_{AH}}}^\text{\eqref{eq:arrhenius}}$& \SI{40667.275}{\kelvin}&
			${k_\mathrm{T_{ref}}}^\text{\eqref{eq:arrhenius}}$& \SI{298.92} {\kelvin}&
			${k_\mathrm{T_L}}^\text{\eqref{eq:arrhenius}}$& \SI{285}{\kelvin}\\
			${k_\mathrm{T_H}}^\text{\eqref{eq:arrhenius}}$ & \SI{308.96}{\kelvin}&
			${k_\mathrm{r_{max}T}}^\text{\eqref{eq:logan-10mod}}$& $\SI{1.0}{}^*$ &
			${k_\mathrm{r_{base}T}}^\text{\eqref{eq:logan-10mod}}$ & $\SI{0.215}{}^*$\\
			${k_\mathrm{\rho T}}^\text{\eqref{eq:logan-10mod}}$& \SI{0.2487}{\per\celsius\per\day}&
			${k_\mathrm{T_{max}}}^\text{\eqref{eq:logan-10mod}}$ & \SI{39.769}{\celsius} &                       
			${k_\mathrm{\Delta T}}^\text{\eqref{eq:logan-10mod}}$& \SI{3.0}{\celsius}\\
			${k_\mathrm{B_{half}dm}}^\text{\eqref{eq:feed_monod}}$& \SI{0.0049}{\gram\per\day}&
			${k_\mathrm{r_{max}dm}}^\text{\eqref{eq:feed_monod}}$& $\SI{0.9758}{}^*$&
			${k_\mathrm{B_{half}gm}}^\text{\eqref{eq:feed_monod_growth}}$& \SI{0.00532}{\gram\per\day}\\
			${k_\mathrm{r_{max}gm}}^\text{\eqref{eq:feed_monod_growth}}$& $\SI{2}{}^*$&
			${k_\mathrm{W_{med}C1}}^\text{\eqref{eq:r_w}}$ & \SI{0.329}{\kilogram\per\kilogram} &
			${k_\mathrm{W_{med}C2}}^\text{\eqref{eq:r_w}}$ & \SI{0.69}{\kilogram\per\kilogram} \\
			${k_\mathrm{W_{med}C3}}^\text{\eqref{eq:r_w}}$ & \SI{0.76}{\kilogram\per\kilogram} &
			${k_\mathrm{W_{med}crit}}^\text{\eqref{eq:r_w}}$ & \SI{0.833}{\kilogram\per\kilogram} &
			${k_\mathrm{r_{max}A}}^\text{\eqref{eq:r_O_monod}}$ & $\SI{1.128}{}^*$ \\
			${k_\mathrm{A_{half}}}^\text{\eqref{eq:r_O_monod}}$ & \SI{0.1877}{\litre\per\minute\per\kilogram}&
			${k_\mathrm{B_{half}dm}}^\text{\eqref{eq:feed_monod}}$& $\SI{0.0137}{\gram\per\day}^+$&
			${k_\mathrm{r_{max}dm}}^\text{\eqref{eq:feed_monod}}$& $\SI{1}{}^{*+}$\\
			${k_\mathrm{B_{half}gm}}^\text{\eqref{eq:feed_monod_growth}}$& $\SI{0.0717}{\gram\per\day}^+$&
			${k_\mathrm{r_{max}gm}}^\text{\eqref{eq:feed_monod_growth}}$& $\SI{1}{}^{*+}$&
			${k_\mathrm{\alpha_{excr}}}^\text{\eqref{eq:epsilon_assim}}$  & \SI{0.5762}{}\\
			${k_\mathrm{\alpha_{assim}}}^\text{\eqref{eq:epsilon_assim}}$  & \SI{0.2135}{}&
			${\epsilon_\mathrm{inges}}^\text{\eqref{eq:epsilon_assim}}$  & \SI{0.79}{}&
			${k_\mathrm{inges}}^\text{\eqref{eq:dB_dry_final}}$  	& \SI{1.61e-4}{\gram\per\gram\per\second}\\
			${k_\mathrm{maint}}^\text{\eqref{eq:dB_dry_final}}$  	& \SI{5.6779e-06}{\gram\per\gram\per\second}&
			${k_\mathrm{T_{\Sigma}1}}^\text{\eqref{eq:r_assim}}$  	& \SI{234.35}{\hour}&
			${k_\mathrm{T_{\Sigma}2}}^\text{\eqref{eq:r_assim}}$  	& \SI{265.5}{\hour}\\
			${k_\mathrm{T_{\Sigma}3}}^\text{\eqref{eq:r_assim}}$  	& \SI{297.5}{\hour}&
			${k_\mathrm{B_{asy}}}^\text{\eqref{eq:r_assim}}$  	&  \SI{0.115}{\gram} &&\\
			\thickhline
		\end{tabular}
		\begin{flushleft} \small $^*$~values are normalized.\\
			$^+$~parameters re-estimated for data set D1 and D5.
		\end{flushleft}
		\label{table:parameters}
	\end{adjustwidth}
\end{table}


\section{Conclusion}
Based on comprehensive data sets aggregated from literature, various factors that affect the growth and development of larvae were analysed.
Models were developed based on literature and based on the analysis of data to accurately and plausibly 
describe the influence of different environmental conditions such as temperature, feed density, feed moisture, and airflow rate on growth and development rates of BSF larvae.
Building on the principles of mass balance, von Bertalanffy and DEB models, a novel dynamic model describing the growth of \textit{Hermetia illucens} larvae was developed.
Concept of development sums was proposed to establish a relationship between growth and development.
The comprehensive dynamic model was obtained consisting of two differential equations, larval dry mass $B_\mathrm{dry}$ and development sums $T_\mathrm{\Sigma}$, and combining the different rate equations.
Model parameters were estimated for all the proposed models based on extensive data sets from different literature and the models were validated with $R^2 > 0.90$ with only few exceptions.
The resulting dynamic model describing the growth and development of the \textit{Hermetia illucens} larvae was validated, using parameters obtained from only a subset of data, on all available data sets.
The dynamic model, proposed and validated in this work, consistently explained: the change of larval dry mass over time; transition of development phase and its effect on growth; and influence of external factors on the larval growth and development.
Performance of the model could be further improved with newer and larger data sets that could reveal other mechanisms or biological processes not explored in this work.
Extension of this work with considerations for energy and resource efficient production of \textit{Hermetia illucens} larvae in large scale production environments will be the future goal.


\section*{Acknowledgments}
This research was partially funded by ESF grant number 100316180.


\end{document}